\documentclass[structabstract]{aa}  

\usepackage{natbib}
\usepackage{graphicx}
\usepackage{txfonts}

\def\smallsun{\hbox{$_\odot$}}
\def\degr{\hbox{$^\circ$}} 
\def\arcmin{\hbox{$^\prime$}}
\def\arcsec{\hbox{$^{\prime\prime}$}}
\def\cm3{cm$^{-3}$}

\begin{document} 


\title{Abundances in planetary nebulae: NGC\,1535, NGC\,6629, He2-108, and 
Tc1\thanks{Based on observations with the 
    {\em Spitzer Space Telescope}, which is operated
    by the Jet Propulsion Laboratory, California Institute of Technology.}}

\author{S.R.\,Pottasch\inst{1}, R.\,Surendiranath\inst{2}, \and 
J.\,Bernard-Salas\inst{3,4}}

\offprints{pottasch@astro.rug.nl} 

\institute{Kapteyn Astronomical Institute, P.O. Box 800, NL 9700 AV
Groningen, the Netherlands \and  T-1, 10/5, 2nd Cross,  
29th Main, BTM I Stage, Bangalore-560068, India \thanks{formerly with the
Indian Institute of Astrophysics, Bangalore} \and Institut
d’Astrophysique Spatiale, Paris-Sud 11, 91405 Orsay, France \and Center for Radiophysics and 
Space Research, Cornell University, Ithaca, NY 14853}

\date{Received date /Accepted date}

\abstract
{Models have been made of stars of a given mass which produce planetary nebulae which usually begin on the AGB (although they may begin earlier) and run to 
the white dwarf stage. While these models cover the so-called dredge-up phases 
when nuclear reactions occur and the newly formed products are brought to the
surface, it is important to compare the abundances predicted by 
the models with the abundances actually observed in PNe.}
{The aim of the paper is to determine the abundances in a group of PNe with
uniform morphological and kinematic properties. The PNe we discuss are circular
with rather low-temperature central stars and are rather far from the galactic
plane. We discuss the effect these abundances have on determining the evolution
of the central stars of these PNe.}
{The mid-infrared spectra of the planetary nebulae \object{NGC\,1535},
\object{NGC\,6629}, \object{He2-108}, and \object{Tc1} (\object{IC\,1266}) 
taken with the {\em Spitzer Space Telescope} are presented. These 
spectra were combined with the ultraviolet {\em IUE} spectra and with the 
spectra in the visual wavelength region to obtain complete, extinction-corrected spectra.  The chemical composition of these nebulae is then found
by directly calculating and adding individual ion abundances. For two of these
PNe, we attempted to reproduce the observed spectrum by making a model nebula. 
This proved impossible for one of the nebulae and the reason for this is 
discussed. The resulting abundances are more accurate than earlier studies
for several reasons, the most important is that inclusion of the far infrared
spectra increases the number of observed ions and makes it possible to include 
the nebular temperature gradient in the abundance calculations.}
{The abundances of the above four PNe have been determined and compared to the
abundances found in five other PNe with similar properties 
studied earlier. These abundances are further compared with values predicted by
the models of Karakas (2003). From this comparison we conclude that
the central stars of these PNe originally had a low mass, probably between
1M\smallsun~ and 2.5M\smallsun. A further comparison is made with the stellar 
evolution models on the HR diagram, from which we conclude that the core mass 
of these PNe is between 0.56M\smallsun~and 0.63M\smallsun.}
{A consistent picture of the evolution of this group of PNe is found that 
agrees with the predictions of the models concerning the present nebular
abundances, the individual masses, and luminosities of these PNe. The distance 
of these PNe can be determined as well.}


\authorrunning{Pottasch et al.}
\titlerunning{Abundances in NGC\,1535 et al.}  

\maketitle

\section{Introduction}

The evolution of planetary nebulae (PNe) has been studied for about four
decades with the result that a general picture of the evolution is understood,
but the details are still being debated. This is both because many of the 
parameters involved are poorly known and because the atmosphere of the central
star is difficult to describe, including the effective temperature, radius, 
and mass loss rate. The (in many cases) uncertain distance enhances these 
difficulties. Furthermore, the interaction of the central star with the nebula 
is often unclear. This manifests itself in a difficulty explaining the 
ionization structure in some PNe.

The abundances of some elements change in the course of evolution of the 
central star. We discuss here primarily the elements helium, nitrogen, carbon, 
and oxygen, although neon, argon, sulfur, chlorine, and sometimes other 
elements are determined as well. These elements are ejected by the star and 
are measured in the nebula. These abundances provide complementary information 
which aid in understanding the evolution and which can eventually confirm or 
limit the results of models of the evolution.

We are concerned here with abundances found in PNe whose central stars are
rather bright and whose morphology is usually described as round or elliptical.
We present new abundances for four such nebulae: NGC1535, NGC6629, Tc1 (IC\,1266), and He2-108. The abundances in these nebulae will be combined with the 
results found earlier in five other PNe, which are also excited by bright 
stars and which have similar (round or elliptical) morphology. These PNe have 
similar spatial properties as well: they are at rather high galactic latitudes.
We can therefore speak of a class of nebulae. This class has been extensively 
studied earlier but from a different point of view. The central stars of these 
PNe are among the brightest known in the visible. It is therefore possible to 
observe these spectra with very high resolution. The profiles of hydrogen and 
helium lines can then be analyzed with the initial goal of obtaining the 
effective temperature (T$_{eff}$) and gravity of these stars. The final goal is
to use these quantities to study the mass and luminosity (thus the evolution) 
of these stars, but additional information is required to do this.

One of the early extensive studies of the line profiles is that of \cite{mendez,mendez2}. These authors take high resolution optical spectra of 
24 PNe and interpret these spectra using NLTE plane-parallel static model 
atmospheres which contain only hydrogen and helium. The values of T$_{eff}$ and
the stellar helium abundance are determined by fitting the observed HeI and 
HeII line profiles while the gravity (log g) is found by fitting the hydrogen 
line profiles (usually H$\gamma$ but sometimes H$\beta$). Then, making use of 
theoretical evolution diagrams which plot T$_{eff}$ against log g for 
different values of the stellar mass M$_{s}$/M\smallsun, ~the value of 
M$_{s}$/M\smallsun~is determined. Using this mass and the value of gravity, 
the angular radius of the exciting
star is determined. Combining this value with the previously found temperature 
and the observed magnitude corrected for extinction, the distance to the 
nebulae is found. In the course of time this group has improved the model 
atmospheres used by first including the sphericity of the atmosphere \citep{kudritzki1} and then by including other elements besides H and He 
in the atmosphere \citep{kudritzki2}. These improved models do
not change the stellar masses or PNe distances substantially. Both the masses
and the distances determined in this way, however, are suspect. Quoting
\cite{kudritzki2} `the masses determined seem systematically 
larger than white dwarf masses and some of the objects ... have unrealistically
high masses'. Concerning the distances: for the 18 cases where the distance 
thus found may be compared with statistical distance \citep{stang2,cahn} they are always higher, in 11 cases more 
than 60\% higher.

Recently \cite{pauldrach} used a different approach. By 
including models losing mass they are able to use 'observed' mass loss rates
and terminal velocities to obtain the stellar masses and distances. These 
masses and distances are even higher than found by \cite{kudritzki2} and are shown to be implausibly high by \cite{napiwotzki}. Two
arguments are used. Because high mass central stars are probably descended from
high mass progenitors they should have the properties of high mass stars.
First they should belong to the disk population of the galaxy which is 
characterized by small scale heights perpendicular to the galactic disk. 
Secondly such high mass star should have produced and dredged up substantial
amounts of both nitrogen and helium. Napiwotzki concludes that on both of these
points the high masses found by \cite{pauldrach} are unlikely.

This suggests the following step: the abundances of those elements produced by
the various stars should be accurately determined. These abundances may then be
used to determine the mass of the star in question by comparing the observed 
abundance with the abundances determined by nucleosynthesis model calculations 
made in the course of evolution of stars of different masses. For these models,
the calculations made by \cite{karakas} are used. These masses can 
then be compared with those found by \cite{kudritzki2}. 
Because the masses used in the
calculations of Karakas are the initial stellar masses, and the masses given
by \cite{kudritzki2} are the present (almost final) stellar
masses, a ratio between the initial and final mass must be known. For this 
secondary information the values given by \cite{weide} will be used.
Unfortunately the uncertainties present in each of these steps will accumulate
so that definite conclusions are difficult to draw. It is however important 
that central star masses can be derived from accurate nebular abundances.

The main purpose of this paper is to obtain accurate abundances for four of the
nebulae discussed by \cite{kudritzki}.  The most important
reason that this can be achieved is the inclusion of the mid-infrared spectrum 
taken with the IRS spectrograph of the {\em Spitzer Space Telescope} (Werner et
al. 2004). The reasons for this have been discussed in earlier papers
\citep[e.g. see][]{pott1,pott2, pott4, bernard}, and can be summarized as follows: 1) The
intensity of the infrared lines is not very sensitive to the electron
temperature nor to possible extinction effects.  2) Use of the
infrared line intensities enable a more accurate determination of the
electron temperature for use with the visual and ultraviolet lines.
3) The number of observed ionization stages is doubled.

For two of the PNe a second method of determining the abundances is attempted
using a nebular model. This has several advantages. First it
provides a physical basis for the electron temperature determination.
Secondly it permits abundance determination for elements which are
observed in only one, or a limited number of ionic stages. This is
true of Mg, Fe, P and Cl, which would be less reliably determined
without a model. A further advantage of modeling is that it provides
information on the central star and other properties of the nebula.

A disadvantage of modeling is that there are more unknowns than
observations and some assumptions must be made especially concerning the 
geometry and the form of the radiation field of the exciting star. The 
difficulties we encountered in determining a model for NGC\,1535 are discussed
below.

Abundance determination for these nebulae have been made earlier and will be
compared with our values in the individual sections.

This paper is structured as follows. In Sect.\, 2, 3, 4 and 5 the observations
of NGC\,1535, NGC\,6629, Tc1 and He2-108 will be discussed and the resultant
abundances will be given and compared to earlier results. In Sect.\,7, the
models for NGC\,1535 and Tc1 are presented and discussed.  In Sect.\,8 the 
evolutionary state of the nebulae is discussed.  Finally, our conclusions
are given in Sect.\,9.

\section{NGC\,1535}

NGC\,1535 (PN G206.4-40.5) is a roughly circular nebula consisting of a bright
inner ring of about 20\arcsec x 17\arcsec surrounded by a much fainter outer
shell of 48\arcsec x 42\arcsec \citep{banerjee}. \cite{tylenda} list a size of 33\arcsec x 32\arcsec down to the 10\%
level. The nebula is at a very high galactic latitude. The distance to the 
nebula is rather uncertain. \cite{ciar} give a value of 2.3 kpc
on the basis of a possible association of a nearby star with the PN. \cite{herald} use a value of 1.6 kpc on the basis of nebular models.
We shall use the latter value when necessary for two reasons. First we feel 
that the evidence for association of the star and nebula is rather weak. 
Secondly at the larger distance the dimensions of the nebula are rather large
for such a bright nebula. We stress however that the distance is uncertain.

\subsection{The infrared spectrum}

Observations of NGC\,1535 were made using the Infrared Spectrograph
\citep[IRS][]{houck} on board the {\em Spitzer Space
 Telescope} with AOR keys of 4111616 (on target) and 4111872
(background). The reduction started from the {\em droop} images
which are equivalent to the most commonly used Basic Calibrated Data
({\em bcd}) images but lack stray-cross removal and flat-field.  The
data were processed using the s15.3 version of the pipeline and
using a script version of {\em Smart} \citep{higdon}.
The tool {\em irsclean} was used to remove rogue pixels. The
different cycles for a given module were combined to enhance the S/N. Then the
resulting high-resolution modules (HR) were extracted using full aperture 
measurements, and the SL measurements setting a window column extraction. This 
same reduction has been used for all nebulae discussed in this paper.

The IRS high resolution spectra have a spectral resolution of about
600, which is a factor of between 2 and 5 less than the resolution
of the ISO SWS spectra.  The mid-infrared measurements are made with
several different diaphragm sizes.  Because the diaphragms
are smaller than the size of the nebulae and are all of differing size,
we first discuss how the different spectra are placed on a common
scale.


Two of the three diaphragms used have high resolution: the short high module 
(SH) measures from 9.9$\mu$m to
19.6$\mu$m and the long high module (LH) from 18.7$\mu$m to
37.2$\mu$m. The SH has a diaphragm size of 4.7\arcsec x 11.3\arcsec,
while the LH is 11.1\arcsec x 22.3\arcsec. If the nebulae are
uniformly illuminating then the ratio of the intensities would
simply be the ratio of the areas measured by the two diaphragms.
Since this is not so, we may use the ratio of the continuum intensity 
in the region of wavelength overlap at 19$\mu$m. These continua are equal 
when the SH intensities are increased by a factor of 3.2. The third
diaphragm is a long slit which is
4\arcsec~wide and extends over the entire nebula. This SL module measures in
low resolution and measures between 5.5$\mu$m and 14$\mu$m. These spectra
are normalized by making the lines in common between the SL and SH
modules to agree. Especially important is the agreement of the
\ion{[S}{iv]} line at 10.51$\mu$m. A uniform scale is obtained by increasing
the SH intensities by a factor of 3.2 with respect to the LH intensities. The
SL intensities are increased by a factor of 1.23. 

The IRS measurement of NGC\,1535 was centered at RA(2000)
04$^{h}$14$^{m}$15.9$^{s}$ and Dec(2000)
-12\degr12\arcmin21\arcsec. This is almost exactly the same as the
value measured by \cite{kerber} of RA(2000)
04$^{h}$14$^{m}$15.78$^{s}$ and Dec(2000)
-12\degr44\arcmin21.7\arcsec, which is presumably the coordinate of
the central star. Thus the IRS measurement was well centered on the
nebula.  The fluxes were measured using the Gaussian
line-fitting routine. The measured emission line intensities are given in 
Table 1, after correcting the SH measurements by the factor 3.2 and the SL
measurements by a factor of 1.23, in the 
column labeled `intensity'. The H$\beta$ flux found from the infrared
hydrogen lines (especially the lines at 12.37$\mu$m) using the theoretical
ratios of \cite{hummer},
is 2.3x10$^{-11}$erg~cm$^{-2}$~s$^{-1}$, which is about 49\% of the total
H$\beta$ intensity. This is reasonable since the well-centered LH diaphragm 
covers a large fraction of the nebula. Note that by scaling in this way the 
nebula is assumed homogeneous when it is larger than the different modules.

\begin{table*}[htbp]
\caption[]{IRS spectra of the four nebulae. The measured line intensity is
 given in Col.3, 5, 7 and 9. Col. 4, 6, 8 and 10 give the ratio of the line
 intensity to H$\beta$(=100).}

\begin{center}
\begin{tabular}{|l|r|cc|cc|cc|cc|}
\hline
\hline
 &   & NGC1535 & NGC1535 & Tc1 & Tc1 & He2-108 & He2-108 & NGC6629 & NGC6629\\
Identification &  $\lambda$($\mu$m) & Intensity$^{\dagger}$ & I/H$\beta$  & Intensity$^{\dagger}$ & I/H$\beta$ & Intensity$^{\dagger}$ & I/H$\beta$ & Intensity$^{\dagger}$ & I/H$\beta$ \\
\hline

  F           &  6.48  &     &     &  198$\pm$27 & 3.2 &     &    &    &   \\
\ion{[Ni}{ii]} ? & 6.63  &     &    &  76.5$\pm$8   & 1.24 &   &     &   &   \\
\ion{[Ar}{ii]} & 7.026 &     &     & 2020$\pm$70 & 32.8 &    &    &     &    \\
\ion{H}{i} (6-5) & 7.47 & 36.2$\pm$2.1  &    & 230$\pm$17 &  & 40.7$\pm$4.8 &    & 306$\pm$29  &    \\
 F     &  7.757  &      &    &     &    &     &     &   37.2$\pm$13 & 0.44 \\
 F    &   8.497  &     &    & 384$\pm$10 &    &    &    &     &     \\
\ion{[Ar}{iii]} & 8.99 & 76.5$\pm$4.3 & 3.32 & 399$\pm$16 & 6.50 & 299$\pm$13 &
25  & 1080$\pm$45 &   \\
\ion{[S}{iv]} & 10.512 & 915$\pm$35 & 39.8 & 29.2$\pm$2.2 & 0.47 & 41.0$\pm$2.4 & 3.6 &1220$\pm$46 &  14.4 \\
\ion{H}{i} (9-7) & 11.305 &   &   & 20.6$\pm$3 &   &   &   &  32.4$\pm$2.3 & \\
\ion{[Cl}{iv]} & 11.763 & 10.5$\pm$1.9 & 0.455  &   &   &   &    &    &  \\
\ion{H}{i} (7-6+11-8) & 12.375 & 22.4:$\pm$2.1 &  & 67.9$\pm$3.6 &  & 11.7$\pm$1.2 &    & 95.8$\pm$4.3 &    \\
\ion{[Ne}{ii]} & 12.81 & 9.55$\pm$1.9  & 0.415 & 2301$\pm$200 & 37.5 & 1175$\pm$130 & 102 & 839$\pm$19 & 102 \\
\ion{[Ne}{iii]} & 15.555 & 1920$\pm$40 & 83.5 & 89.5$\pm$5 & 1.46 & 189$\pm$6 & 16.5 & 7530$\pm$ & 89.1 \\
F   & 17.66 &     &     &     &    & 35$\pm$12 & 3 &   &   \\
\ion{[P}{iii]} & 17.88 &   &    &   &    &    &    & 51$\pm$2.3 & 0.61 \\
\ion{[S}{iii]} & 18.714 & 67.4$\pm$3.3  & 2.93 & 863$\pm$80 & 14.0 & 830$\pm$150 & 72.2 & 1540$\pm$32 & 18.2\\
\ion{[Cl}{iv]} & 20.316 & 10.8 $\pm$0.9  & 0.47 &   &   &   &   &   & \\
\ion{[Ar}{iii]} & 21.820 & 7.44 $\pm$0.73 & 0.323 & 26.6$\pm$2 & 0.432 & 19.4$\pm$2 & 1.69 & 85.4$\pm$6.9 & 1.01 \\
\ion{[Fe}{iii]} & 22.92 &    &    & 18.1$\pm$1.6 & 0.295 & 11.7$\pm$0.7 &1.02 &  &   \\
\ion{[O}{iv]} & 25.889 & 2466 $\pm$21  & 107 &   &   &    &    &   &  \\
\ion{[S}{iii]} & 33.48 & 63.2 $\pm$1.9  & 2.75 & 382$\pm$33 & 6.21 & 402$\pm$18 & 35.0 & 807$\pm$34 & 6.3 \\
\ion{[Ne}{iii]} & 36.013 & 203$\pm$2.2 & 8.84  &   &   &   &   &  531$\pm$24 & 6.3 \\

\hline

\end{tabular}
\end{center}

$^{\dagger}$ Intensities measured in units of 10$^{-14}$erg~cm$^{-2}$~s$^{-1}$.
The intensities SH measurements (below 19$\mu$m) and the SL measurements have 
been increased by a factor to bring them all to the scale of the LH 
measurements. The factors used are given in text. Lines identified by 'F' are 
related to the Fullerine molecule \citep{cami}. \\

\end{table*}

\subsection{The visual spectrum}

The visual spectrum has been measured by at least six authors. We list here
the results from four of these. The line intensities listed have been corrected
by each author for a value of extinction determined by them to obtain a
theoretically correct Balmer decrement. The result are listed in Table 2, where
the last column lists the average value which we have used. No attempt has been
made to use a common extinction correction because then the Balmer decrement 
will be incorrect. The value of extinction C which the individual authors found
is listed at the bottom of the table. None of the spectra measure the weaker 
lines very well. The errors may be judged by the agreement (or disagreement) 
of the various measures and appear to be within 20\% for the stronger lines 
and worse for the weaker lines.

\begin{table}[h]
\caption[]{Visual spectrum of NGC\,1535.}
\begin{center}
\begin{tabular}{llccccc}
\hline
\hline
\multicolumn{1}{c}{$\lambda$} & Ion & \multicolumn{4}{c}{Intensities$^{\dagger}$}& Average\\ \cline{3-6}
\multicolumn{1}{c}{(\AA)}& & (1) & (2)& (3) & (4) & Intens.\\
\hline

3727$^{\ast}$   & \ion{[O}{ii]} & 4.1 & 7.1    & 8.41  & 9.37 & 8.41 \\
3869 & \ion{[Ne}{iii]} & 90 & 95.8    & 97  & 116 &  101 \\
4267 & \ion{C}{ii}    & 0.4:   & 0.34   & 1.03  & 0.36  & 0.34   \\
4363 & \ion{[O}{iii]} & 13.3 & 12.5   &  11  & 12.6 &  12.6 \\
4686 & \ion{He}{ii}  & 27 & 14.2 & 18.3  & 17 & 17.5 \\
4711 & \ion{[Ar}{iv]} & 4.5  & 4.2     & 5.4  & 4.5 & 4.5  \\
4740 & \ion{[Ar}{iv]} & 3.8    & 3.2  & 3.8   & 3.36 & 3.4  \\
4861 & \ion{H$\beta$} & 100 & 100 & 100 & 100 & 100 \\
5007 & \ion{[O}{iii]} & 1220 & 1210 & 1180 & 1190 & 1200 \\
5517 & \ion{[Cl}{iii]} &     & 0.242  & 0.264  &    & 0.25   \\
5538 & \ion{[Cl}{iii]} &     & 0.177  & 0.175  &   & 0.175 \\
5755 & \ion{[N}{ii]}  &      & 0.1:  & 0.62  &    & 0.1:   \\
5876 & \ion{He}{i}    & 10.2   & 12.6 & 11.8  & 12.6 & 12.2 \\
6312 & \ion{[S}{iii]} &       & 0.303   &      & 0.28   & 0.30 \\
6584 & \ion{[N}{ii]}  & 1.8  & 1.62  & 0.25:   & 0.85 & 1.6: \\
6717 & \ion{[S}{ii]}  &      & 0.0627  &      &     & 0.0627  \\
6731 & \ion{[S}{ii]}  &      & 0.114  &      &     & 0.114   \\
7135 & \ion{[Ar}{iii]}& 6.0 & 6.42   & 5.3  &     & 6.3  \\
7263 & \ion{[Ar}{iv]}  &      & 0.217  &      &     & 0.217   \\
8045 & \ion{[Cl}{iv]}  &        & 0.512  & 0.53     &      & 0.52   \\
9532 & \ion{[S}{iii]}  &     & 6.75  &     &     & 6.75 \\
C(H$\beta$)    &        & 0.2 & 0.07  & 0.11  & 0.01 &   \\
\hline
\end{tabular}
\end{center}

$^{\dagger}$ References; (1) \cite{barker}, (2)\cite{milingo1}, (3) \cite{aller}, (4) \cite{krabbe}.\\ 
(:) indicates uncertain values.\\ 
$^{\ast}$ This is a blend of $\lambda$3726 and $\lambda$3729 lines. Only Aller 
\& Czyzak are able to resolve this doublet: 3726=5.4, 3729=3.01\\
C is the extinction used by the author
\end{table}

\subsection{The ultraviolet spectrum }

Quite a large number of IUE spectra of this nebula have been taken. If only 
those taken with the large diaphragm are counted there are six shortwavelength 
spectra and five longwavelength spectra with the central star included. These
have been made with low resolution. Also two shortwavelength spectra (SWP 10821
and 13495) and one longwavelength spectrum (LWR02165) were taken with high 
resolution. They also included the central star. Two spectra are also available
which included only the nebula (SWP15497 and LWR11975). These spectra are 
especially useful because longer exposures can be made without saturating the
stronger lines. The \ion{N}{iii} line can only be seen on these spectra. The 
IUE diaphragm is an ellipse about 10\arcsec x 21\arcsec; the wavelength range 
is from 1150\AA~to about 3220\AA. The high resolution spectra have a resolution
of 0.2\AA~while the low resolution is about 6\AA. There are also two spectra 
taken with the Hopkins Ultraviolet Telescope (HUT), one taken with the central 
star in the diaphragm, the other has only the nebula in the 9.4\arcsec x 116\arcsec 
diaphragm. The HUT spectra have a resolution of about 3\AA~and they cover the
wavelength range between 830\AA~and 1860\AA.

Because the nebula is in all cases larger than the diaphragm used and the 
various spectra are centered at different positions in the nebula, a total 
spectrum is obtained by normalizing the individual spectra to the strong 
\ion{He}{II} line at 1640\AA. Thus as basis we use the high resolution IUE 
spectra SWP13495 and LWR02165; the other lines are found by using the measured
ratio to the intensity of the 1640\AA~line in the individual spectra. The 
values are given in the third column of Table 3. The measured values are then
corrected for the diaphragm size and the extinction using the total H$\beta$
flux of 4.73 x 10$^{-11}$ erg cm$^{-2}$ s$^{-1}$ (see below), then the intensity of the
\ion{He}{ii} at 4686\AA~becomes 8.25 x 10$^{-11}$ erg cm$^{-2}$ (from Table 2).
The theoretical helium spectrum \citep{hummer} then gives
the intensity of the 1640\AA~line to be 5.64 x 10$^{-11}$ erg cm$^{-2}$. 
The other lines are then corrected for their extinction with respect to the
1640\AA~line using the values given by \cite{fluks} and the value 
C=0.09 as given below. These values are listed in col.4 of Table 3. In col.5
of the table the ratio of the line to H$\beta$ normalized to H$\beta$=100 is 
given. The uncertainties in the intensities are estimated to be 20\% for the 
stronger lines and 30\% for the weaker lines.

\begin{table}[htbp]
\caption[]{UV Spectrum of NGC\,1535.}  
\begin{center}
\begin{tabular}{llccc}
\hline
\hline
\multicolumn{1}{c}{$\lambda$} & Ion &\multicolumn{3}{c}{Intensities}\\
\cline{3-5}
\multicolumn{1}{c}{(\AA)}& & (1) & (2) &  (I/H$\beta$)   \\
\hline

977  & \ion{C}{iii]} & 9.6 & 8.4 & 18 \\
1175 & \ion{C}{iii]} & 34  & 23.5 & 50 \\
1548 & \ion{C}{iv}  &  30.8  & 19.3  & 40.7  \\
1550 & \ion{C}{iv}  &  16.8  & 10.0  & 21.2  \\
1640 & \ion{He}{ii}  & 95   & 56.4  & 119  \\
1661 & \ion{O}{iii]} & 1.8   & 1.1 & 2.3 \\
1663 & \ion{O}{iii]} & 6.9   & 4.1 & 8.7 \\
1678 &   ?          &  4.8   & 2.8  & 6.0 \\
1750 & \ion{N}{iii]} & 8.5  & 5.05 & 10.7 \\
1906 & \ion{C}{iii]} & 85.2    & 50.9  & 108  \\
1909 & \ion{C}{iii]} & 58.0    & 34.5  & 73  \\
2297 & \ion{C}{iii]} & 16    & 9.5   & 20  \\
2734 & \ion{He}{ii}  & 3.5  & 1.9 & 4.0  \\
2837 & [\ion{Fe}{iv}]?  & 1.8 & 1.0 & 2.1  \\
3048 & \ion{O}{iii}  & 7.0   & 3.6   & 7.2   \\
3134 & \ion{O}{iii}  & 28.0  & 14.2  & 30   \\
3204 & \ion{He}{ii}  & 7.2  & 3.7 & 7.7 \\

\hline 
\end{tabular} 
\end{center}
(1)Measured intensity from in units of 10$^{-13}$ erg cm$^{-2}$ s$^{-1}$. \\
(2)Intensity corrected for diaphragm size and extinction in units of
10$^{-12}$ erg cm$^{-2}$ s$^{-1}$. 
I/H$\beta$ is normalized to H$\beta$=100 \\ 

\end{table}

\subsection{Extinction}

The two methods which can be used for obtaining the extinction are:
(1) comparison of radio emission with H$\beta$ flux, and (2)
comparison of observed and theoretical Balmer decrement. The four
values of the extinction correction C(H$\beta$) which are found in the
literature are given in Table 2, and are seen to have a rather large
range.  Let us discuss the radio emission and the H$\beta$ flux.

\subsubsection{The 6\,cm radio emission and the H$\beta$ flux }

The 6\,cm flux density has been measured by \cite{griffith} 
as 168 mJy. This corresponds to an H$\beta$ flux of 4.73 x 10$^{-11}$ erg 
cm$^{-2}$ s$^{-1}$ using the electron temperature and helium abundance given
below. The observed H$\beta$ flux of 3.8 x 10$^{-11}$ erg cm$^{-2}$ s$^{-1}$ 
\citep[see][]{cahn}. This leads to  an extinction constant of 
C(H$\beta$)=0.094 or $E_{\mathrm{B-V}}$=0.064. This is quite similar to the 
values found from the Balmer decrement and listed in Table 2. This value,  
together with the extinction curve of \cite{fluks}. has
been used in correcting the UV fluxes in Table 3.

\section{Chemical composition of NGC\,1535}

The method of analysis is the same as used in the papers cited in the
introduction. First the electron density and temperature as function
of the ionization potential are determined. Then the ionic abundances
are determined, using density and temperature appropriate for the ion
under consideration, together with Eq.(1). Then the element abundances
are found for those elements in which a sufficient number of ion
abundances have been derived.

\subsection{Electron density}

The ions used to determine $N_{\mathrm{e}}$ are listed in the first
column of Table 4. The ionization potential required to reach that
ionization stage, and the wavelengths of the lines used, are given in
Cols. 2 and 3 of the table. Note that the wavelength units are
\AA~when 4 ciphers are given and microns when 3 ciphers are shown. The
observed ratio of the lines is given in the fourth column; the
corresponding $N_{\mathrm{e}}$ is given in the fifth column. The
temperature used is discussed in the following section, but is
unimportant since these line ratios are essentially determined by the
density.

The electron density appears to be about 1000 cm$^{-3}$ although the two ions 
with the lowest ionization potential give a somewhat higher value. These values
are less well determined because the ratios are poorly measured. The density is
probably not uniform as indicated by the structures seen in the central area of
the nebula, which may contribute to this difference. A density of 1000
cm$^{-3}$ is used in the abundance determination in Table 6, but none of the
abundances listed in the table is sensitive to the density in the range shown
in the table.

\begin{table*}[t]
\caption[]{ Observed Electron density indicators in the nebulae.}
\begin{center}
\begin{tabular}{|l|r|r|cc|cc|cc|cc|}
\hline
\hline
Ion &Ioniz. & Lines& Obs.Ratio &N$_{\mathrm{e}}$  (cm$^{-3}$ & Obs.Ratio  &N$_{\mathrm{e}}$  (cm$^{-3}$ & Obs.Ratio &N$_{\mathrm{e}}$  (cm$^{-3}$ & Obs.Ratio &N$_{\mathrm{e}}$  (cm$^{-3}$     \\
&Pot.(eV) & Used  & NGC1535 & NGC1535 & Tc1 & Tc1 & He2-108 & He2-108 & NGC6629 & NGC6629 \\
\hline
\ion{[S}{ii]} & 10.4 & 6731/6716 & 1.8: & 2500: & 1.59  & 2900: &       &       & 0.74  & 1600 \\
\ion{[O}{ii]} & 13.6 & 3726/3729 & 1.8: & 3000: & 1.51  & 1900: &       &       & 1.63  & 2400 \\
\ion{[S}{iii]} & 23.3 & 33.5/18.7 & 0.94 & 700  & 0.444 & 3200  & 0.485 & 2200  & 0.525 & 2200 \\
\ion{[Cl}{iii]} & 23.8 & 5538/5518 & 0.70: & 600: & 1.06 & 3000: &      &       & 0.9:  & 1400: \\
\ion{[C}{iii]} & 24.4 & 1906/1909 & 1.48 & 900  & 1.24  &4500  &        &       &       &      \\
\ion{[Ar}{iv]} & 40.7 & 4740/4711 & 0.76: & 1000: &     &      &        &       &       &      \\
\hline
\end{tabular}  
\end{center}
(:) indicates uncertain values
\end{table*}

\subsection{Electron temperature}

A number of ions have lines originating from energy levels far enough
apart that their ratio is sensitive to the electron temperature. These
are listed in Table 5, which is arranged similarly to the previous
table. While there is a slight scatter in these values there is no clear
indication of a temperature gradient as function of the ionization potential as
has been seen in some other nebulae. An electron temperature of 12\,000 K will
be used with an uncertainty of less than 1\,000 K.


\begin{table*}[t]
\caption[]{Observed Electron temperature indicators in the nebulae.}
\begin{center}
\begin{tabular}{|l|r|r|cc|cc|cc|cc|}
\hline
\hline
Ion & Ioniz. & Lines& Obs.Ratio & $T_{\mathrm{e}}$(K)& Obs.Ratio & $T_{\mathrm{e}}$(K)& Obs.Ratio & $T_{\mathrm{e}}$(K)& Obs.Ratio & $T_{\mathrm{e}}$(K) \\
& Pot.(eV)& Used  & NGC1535 & NGC1535 & Tc1  & Tc1  & He2-108 & He2-108 & NGC6629 & NGC6629  \\
\hline

\ion{[N}{ii]}  & 14.5 & 5577/6584 &       &        & 0.0114 & 9\,300 &        &        &         &       \\
\ion{[S}{iii]} & 23.3 & 6312/18.7 & 0.102 & 13\,000 & 0.0337 & 9\,100 &       &        & 0.029:  & 8\,700: \\
\ion{[Ar}{iii]} & 27.6 & 7135/8.99 & 1.9 & 14\,000  & 0.888  & 9\,000 & 0.60  & 8\,000 & 1.0     & 8\,500 \\
\ion{[Ar}{iii]} & 27.6 & 7135/5192 &     &          & 190    & 9\,000 &       &        &         &        \\
\ion{[O}{iii]} & 35.1 & 4363/5007 & 0.0105 & 11\,800 & 0.00446 & 9\,000 & 0.0055 & 9\,500 & 0.0041 & 8\,700 \\
\ion{[O}{iii]} & 35.1 & 1663/5007  & 0.0092 & 10\,700 &        &        &        &       &       &         \\
\ion{[Ne}{iii]} & 41.0 & 3868/15.5 & 1.21 & 12\,200  &        &         & 0.285  & 8\,200 & 0.42 & 8\,900 \\

\hline
\end{tabular}
\end{center}
\end{table*}

\subsection{Ionic and element abundances}

The ionic abundances have been determined using the following equation:

\begin{equation}
\frac{N_{\mathrm{ion}}}{N_{\mathrm{p}}}= \frac{I_{\mathrm{ion}}}{I_{\mathrm{H_{\beta}}
}} N_{\mathrm{e}}
\frac{\lambda_{\mathrm{ul}}}{\lambda_{\mathrm{H_{\beta}}}} \frac{\alpha_{\mathrm{H_{\beta}}}}{A_{\mathrm{ul}}}
\left( \frac{N_{\mathrm{u}}}{N_{\mathrm{ion}}} \right)^{-1} 
\label{eq_abun}
\end{equation}

where $I_{\mathrm{ion}}$/$I_{\mathrm{H_{\beta}}}$ is the measured
intensity of the ionic line compared to H$\beta$, $N_{\mathrm{p}}$ is
the density of ionized hydrogen, $\lambda_{\mathrm{ul}}$ is the
wavelength of this line, $\lambda_{\mathrm{H_\beta}}$ is the
wavelength of H$\beta$, ${\alpha_{\mathrm{H_\beta}}}$ is the effective
recombination coefficient for H$\beta$, $A_{\mathrm{ul}}$ is the
Einstein spontaneous transition rate for the line, and
$N_{\mathrm{u}}$/$N_{\mathrm{ion}}$ is the ratio of the population of
the level from which the line originates to the total population of
the ion. This ratio has been determined usually using a five level atom. 
Sometimes a two level atom was sufficient.

The results are given in Table 6, where the first column lists the ion
concerned, and the second column the line used for the abundance
determination. The third column gives the intensity of the line used
relative to H$\beta$=100. The fourth column shows the ionic abundances, and the
fifth column gives the Ionization Correction Factor (ICF). This has
been determined empirically, usually by looking at the ionization potential
of the missing ion.  Notice that the ICF is unity for all elements except for
Ar, S and Cl where it is close to unity. The helium abundance has been derived 
with the help of the theoretical work of \cite{benjamin} and 
\cite{porter}.

\begin{table}[htbp] 
\caption[]
{Ionic concentrations and chemical abundances in NGC\,1535.
Wavelength in Angstrom for all values of $\lambda$ above 1000, otherwise
in $\mu$m.}

\begin{center}

\begin{tabular}{lccccc}
\hline
\hline
Ion & $\lambda$ & Int./H$\beta$ &  $N_{\mathrm{ion}}$/$N_{\mathrm{p}}$ 
& ICF & $N_{\mathrm{el.}}$/$N_{\mathrm{p}}$\\
\hline
He$^{+}$  & 5875 & 12.2 &  0.076 &     &      \\
He$^{++}$  & 4686 & 17.5 &  0.015 & 1 & 0.091 \\
C$^{++}$ & 1909 & 181 & 1.23(-4) &    &     \\
C$^{+3}$ & 1548 & 61.9 &  3.54(-5) & 1 & 1.6(-4) \\
N$^{+}$   & 6584 & 1.6 &  2.0(-7) &     &     \\
N$^{++}$  & 1750 & 10.7 &  3.2(-5) & 1  & 3.2(-5) \\
O$^{+}$  & 3727 & 8.4 &  1.38(-6)  &      &     \\
O$^{++}$  & 5007 & 1200 &  2.4(-4) &  &  \\
O$^{+3}$  & 25,9 & 107  &  2.03(-5)  & 1 & 2.7(-4) \\
Ne$^{+}$  & 12.8 & 0.42  &  5.2(-7) &     &    \\
Ne$^{++}$ & 15.5 & 83.5 &  4.70(-5)  &    &    \\ 
Ne$^{++}$ & 3869 & 101 &  4.95(-5)  & 1.1   & 5.4(-5)\\  
S$^{+}$   & 6731 & 0.114 &  3.7(-9)  &     &     \\
S$^{++}$  & 18.7  & 2.93 &  2.29(-7) &    &    \\
S$^{+3}$  & 10.5  & 39.8  &  8.9(-7) & 1.2 & 1.3(-6) \\
Ar$^{++}$ & 8.99  & 3.32 &  2.83(-7) &    &    \\
Ar$^{++}$ & 7135 & 6.3 &  3.57(-7) &      &     \\
Ar$^{+3}$ & 4740 & 3.4  &  5.75(-7) & 1.2  & 1.1(-6)  \\
Cl$^{++}$ & 5538 & 0.175 &  1.5(-8) &       &       \\
Cl$^{+3}$ & 11.8 & 0.455  &  2.24(-8) &      &       \\ 
Cl$^{+3}$ & 8045 & 0.52  &  3.36(-8) & 1.2  & 6.0(-8) \\    
\hline
\end{tabular}
\end{center}
Intensities given with respect to H$\beta$=100.

(:) indicates uncertain values

\end{table}

\subsection{Comparison with other determinations}

In Table 7 the present abundances are compared to earlier determinations. The
agreement is usually within a factor of 2, except for Chlorine which is 
difficult to measure.

\begin{table}[htbp]
\caption[]{Comparison of abundances in \object{NGC\,1535}.}
\begin{center}
\begin{tabular}{lllllll}
\hline
\hline
Ele.  & pres & bark  & milin   &  AC  &  TPP  &  KC    \\ 
\hline  

He     & 0.091  & 0.097 & 0.0.96  & 0.094  & 0.091  & 0.105     \\
C(-4)  &  1.6   & 0.8  &   1.91    & 3.65    &        &        \\
N(-5)  &  3.2   & 4.3  & 2.09      &        &         & 1.3     \\
O(-4)  &  2.7   & 3.3     & 2.98 &  4.05    & 3.8    &   2.6    \\
S(-6)  &  1.3   &       &        &       &        &           \\   
Ar(-6) &  1.1   & 1.2  &  0.937  &  1.93  &       &           \\
Ne(-5) &  5.3   & 7.1   & 6.37   &  8.3  & 9.5 &  7.3    \\
Cl(-7) &  0.60   &      &  0.184  &  1.26  &       &        \\

\hline  

\end{tabular}
\end{center}

References: bark: \cite{barker}, milin: \cite{milingo1} AC: \cite{aller}, TPP: \cite{tpp}, KC: \cite{krabbe}

\end{table}

\section{Tc1 (IC\,1266)}

Tc1 (PN G345.2-08.8, also known as IC\,1266, SaSt\,2-16 and IRAS 17418-4604) is
morphologically quite similar to NGC\,1535. It is roughly circular and has a 
size at the 10\% level of 12.9\arcsec x 12.2\arcsec \citep{tylenda}. A somewhat smaller diameter (9.6\arcsec) is given by \cite{acker}. The size is small enough so that most of the radiation can be measured in 
the IUE diaphragm. The nebula is surrounded by a faint halo which is also 
circular and has a diameter of about 53\arcsec.

The 6cm continuum radio flux density has been measured by \cite{griffith} as 140 mJy. \cite{ma2} have measured the 2cm radio
flux density as 130 mJy, which corresponds to a value of 147 mJy at 6cm. We use
an average value of 145 mJy at 6cm, which corresponds to an H$\beta$ flux of
5.1 x 10$^{-11}$ erg cm$^{-2}$ s$^{-1}$. The measured H$\beta$ flux is
2.18 x 10$^{-11}$ erg cm$^{-2}$ s$^{-1}$ \citep[see][]{acker} which
leads to an extinction coefficient C=0.36.

\subsection{Infrared spectrum}

The IRS measurement of Tc\,1 was centered at RA(2000) 17$^{h}$45$^{m}$35.3$^{s}$ and Dec(2000) -46\degr05\arcmin23.3\arcsec. This is almost exactly the same as
the value measured by \cite{kerber} of RA(2000)
17$^{h}$45$^{m}$35.3$^{s}$ and Dec(2000) -46\degr05\arcmin23.8\arcsec, which is
presumably the coordinate of the central star. Thus the IRS measurement was 
well centered on the nebula.  The measured emission line intensities are given 
in Table 1, after correcting the SH measurements by the factor 2.02 and the SL
measurements by a factor of 2.35, in the column labeled `intensity'. The 
H$\beta$ flux found from the infrared hydrogen lines (especially the lines at 
12.37$\mu$m) using the theoretical ratios of \cite{hummer},
is 6.15x10$^{-11}$erg~cm$^{-2}$~s$^{-1}$, which is about 20\% higher than the 
total H$\beta$ intensity. This indicates that the LH diaphragm measured the 
entire nebula. The measurement of a higher flux in the infrared is within the 
uncertainties of the various measurements. Two of the features in the table 
have been identified as belonging to the fullerene molecules \citep{cami}.

\subsection{Visual spectrum}

There are only a few visual spectra of Tc\,1. The best is the very good 
spectrum reported by \cite{williams}. These authors measured the
nebula at two positions at either side of the central star, but carefully
avoiding the star. They used a rectangular slit 2\arcsec x 4\arcsec. They 
correct their intensities for an extinction found from the Balmer decrement,
These corrected intensities are shown in the third column of Table\,8 for some 
of the lines of interest to us. Notice that the \ion{[Ne}{iii]} line at 
$\lambda$3869 \AA~was too weak to measure. \cite{williams} do 
not report the intensities of any of the \ion{He}{i} lines; for the intensities
of these lines the spectrum reported by \cite{kings} are 
used. These are shown in column 4 of the table, where the average intensity is
weighted to the spectrum of \cite{williams}.

\begin{table}[h]
\caption[]{Visual spectrum of Tc\,1.}
\begin{center}
\begin{tabular}{llccc}
\hline
\hline
\multicolumn{1}{c}{$\lambda$} & Ion & \multicolumn{2}{c}{Intensities$^{\dagger}$}& Average\\ \cline{3-4}
\multicolumn{1}{c}{(\AA)}& & (1) & (2)  & Intens.\\
\hline

3726  & \ion{[O}{ii]} & 130   &     &  130 \\
3729  & \ion{[O}{ii]} & 86   &     &  86 \\
4340 & \ion{H$\gamma$}&      & 44.1 & 44.1    \\
4363 & \ion{[O}{iii]} & 0.555 & 0.45   &  0.55 \\
4471 & \ion{He}{i}    &       & 1.1 & 1.1 \\
4861 & \ion{H$\beta$} & 100 & 100 & 100 \\
5007 & \ion{[O}{iii]} & 124 &    & 124 \\
5192 & \ion{[Ar}{iii]} & 0.0314 &    & 0.314 \\
5517 & \ion{[Cl}{iii]} & 0.285    &     & 0.285    \\
5538 & \ion{[Cl}{iii]} & 0.303   &     & 0.303 \\
5755 & \ion{[N}{ii]}  &  1.09    &      & 1.09   \\
5876 & \ion{He}{i}    &        & 9.01  & 9.01 \\
6312 & \ion{[S}{iii]} &  0.471  &      &   0.471 \\
6584 & \ion{[N}{ii]}  & 95.4  & 95.1  &  95.4 \\
6717 & \ion{[S}{ii]}  &  2.20    &      &  2.20  \\
6731 & \ion{[S}{ii]}  &  3.50    &      &  3.50   \\
7135 & \ion{[Ar}{iii]}& 5.75 & 5.29   &  5.65  \\
9069 & \ion{[S}{iii]}  &  13.0   &     &  13.0 \\
C(H$\beta$)    &        & 0.33 & 0.40  &    \\
\hline
\end{tabular}
\end{center}

$^{\dagger}$ References; (1) \cite{williams}, (2) \cite{kings}.\\ 
(:) indicates uncertain values.\\ 
C is the extinction used by the author
\end{table}

\subsection{The ultraviolet spectrum }

There are 14 IUE spectra of Tc\,1: Three high resolution shortwave spectra, seven low
resolution shortwave spectra and four low resolution longwave spectra. Only a few
lines are strong enough for a good identification as a nebular line however. In
addition the spectrum is of low excitation so that the connection between the
ultraviolet and visual spectra through the \ion{He}{ii} lines of $\lambda$1640 
\AA~and $\lambda$4686 \AA~cannot be made. However the nebula is small enough so
that almost all of its emission is measured. \cite{feibel} has 
reported measuring the \ion{[O}{ii]} line at $\lambda$2471 \AA~which can be 
used to connect the ultraviolet spectrum to the visual spectrum but we feel 
that the spectra are too noisy to measure this line. We find that only two 
lines are clearly measurable on the low resolution spectra. The high resolution
spectra shows more lines but because they may be interstellar or stellar we do
not report them here. Table 10 lists our measurements. The extinction correction
is made using a value of C=0.33 as found by \cite{williams}.

\begin{table}[htbp]
\caption[]{UV Spectrum of Tc\,1.}  
\begin{center}
\begin{tabular}{llccc}
\hline
\hline
\multicolumn{1}{c}{$\lambda$} & Ion &\multicolumn{3}{c}{Intensities}\\
\cline{3-5}
\multicolumn{1}{c}{(\AA)}& & (1) & (2) &  (I/H$\beta$)   \\
\hline

1906 & \ion{C}{iii]} & 1.45    & 7.6  & 14.8  \\
1909 & \ion{C}{iii]} & 1.16    & 6.1  & 11.9  \\
2325 & \ion{C}{ii]} & 3.90    & 23.4   & 45.5  \\

\hline 
\end{tabular} 
\end{center}
(1)Measured intensity from in units of 10$^{-12}$ erg cm$^{-2}$ s$^{-1}$. \\
(2)Intensity corrected for extinction in units of 10$^{-12}$ erg cm$^{-2}$ s$^{-1}$.\\ 
I/H$\beta$ is normalized to H$\beta$=100 \\ 

\end{table}

\subsection{Electron density}

The ions used to determine $N_{\mathrm{e}}$ are listed in the first column of 
Table 4 and in a similar manner as in Table 4. the electron density is given. 
It is about 3000 cm$^{-3}$.

\subsection{Electron temperature}

Five ions have lines originating from energy levels far enough
apart that their ratio is sensitive to the electron temperature. These
are listed in Table 5. An electron temperature of 9\,000 K will be used with 
an uncertainty of less than 500 K.  

\subsection{Ionic and element abundances}

The ionic abundances have been determined using  equation 1 above with an
electron temperature of 9\,000 K and a density of 3\,000 cm$^{-3}$.
The results are given in Table 10, where the first column lists the ion
concerned, the second column the line used for the abundance determination and
the third column gives the intensity of the line used relative to H$\beta$=100.
The fourth column shows the ionic abundances, and the
fifth column gives the Ionization Correction Factor (ICF), determined with the
help of the model described below. In all cases when the ICF is greater than 1,
the principal ionization stage of that element has been observed.

\begin{table}[htbp] 
\caption[]
{Ionic concentrations and chemical abundances in Tc\,1.
Wavelength in Angstrom for all values of $\lambda$ above 1000, otherwise
in $\mu$m.}

\begin{center}

\begin{tabular}{lccccc}
\hline
\hline
Ion & $\lambda$ & Int./H$\beta$ &  $N_{\mathrm{ion}}$/$N_{\mathrm{p}}$ 
& ICF & $N_{\mathrm{el.}}$/$N_{\mathrm{p}}$\\
\hline
He$^{+}$  & 5875 & 9.0 &  0.060 &  ?  & $\geq$0.060    \\
C$^{+}$   & 2325 &  45.5 & 1.85(-4) &    &      \\
C$^{++}$ & 1909 & 26.7 & 1.71(-4) & 1   & 3.6(-4)   \\
N$^{+}$   & 6584 & 95.4 &  2.58(-5) & 1.4    & 3.6(-5)    \\
O$^{+}$  & 3727 & 130 &  1.87(-4)  &      &     \\
O$^{++}$  & 5007 & 124 &  6.8(-5) & 1  & 2.6(-4) \\
Ne$^{+}$  & 12.8 & 37.5  &  6.2(-5) &     &    \\
Ne$^{++}$ & 15.5 & 1.46 &  1.1(-6)  &  1  & 6.3(-5)   \\ 
S$^{+}$   & 6731 & 3.50 &  1.1(-6)  &     &     \\
S$^{++}$  & 6312 & 0.47 &  1.7(-6) &  & \\
S$^{++}$  & 18.7  & 14.0 &  1.58(-6) &    &    \\
S$^{+3}$  & 10.5  & 0.47  &  3.8(-8) & 1 & 2.8(-6) \\
Ar$^{+}$  & 6.99  & 32.8  &  4.2(-6) &    &     \\
Ar$^{++}$ & 8.99  & 6.5 &  7.0(-7) &    &    \\
Ar$^{++}$ & 7135 & 5.75 &  7.0(-7) &  1    & 5.1(-6)   \\
Cl$^{++}$ & 5538 & 0.303 &  5.9(-8) &  1.6  & 9.4(-8)      \\
P$^{++}$ & 17.9 & 0.735  &  1.67(-7) & 1.2  & 2.0(-7)      \\ 
Fe$^{++}$ & 22.9 & 0.295  & 1.1(-7) & 1.4  & 1.54(-7) \\    

\hline
\end{tabular}
\end{center}
Intensities given with respect to H$\beta$=100.

\end{table}

\section{He2-108}

He2-108 (PN G316.1+08.4, also known as IRAS 14147-5156) is morphologically 
quite similar to NGC\,1535 and Tc\,1. It is roughly circular and has a 
size at the 10\% level of 13.6\arcsec x 12.3\arcsec \citep{tylenda}. A somewhat smaller diameter (11\arcsec) is given by \cite{acker}. The size is small enough so that most of the radiation can be measured in 
the IUE diaphragm. 

The 6cm continuum radio flux density has been measured by 
\cite{ma1} as 33 mJy. \cite{ma2} have measured the 2cm radio
flux density as 43 mJy, which corresponds to a value of 49 mJy at 6cm. An 
uncertain average value of 39 mJy at 6cm  corresponds to an H$\beta$ flux of
1.26 x 10$^{-11}$ erg cm$^{-2}$ s$^{-1}$. The measured H$\beta$ flux is
3.7 x 10$^{-12}$ erg cm$^{-2}$ s$^{-1}$ \citep[see][]{acker} which
leads to an extinction coefficient C=0.53.

\subsection{Infrared spectrum}

The IRS measurement of He2-108 was centered at RA(2000) 14$^{h}$18$^{m}$08.4$^{s}$ and Dec(2000) -52\degr10\arcmin38.0\arcsec. This is a slight mispointing 
from the center measured by \cite{kerber} of RA(2000)
14$^{h}$18$^{m}$08.89$^{s}$ and Dec(2000) -52\degr10\arcmin39.7\arcsec, which 
is presumably the coordinate of the central star. This does not have an 
important effect for the LH measurement because of the large LH diaphragm. It 
does however affect the SH measurement for which the Nod 1 measurement measured
only part of the nebula. The Nod 2 measurement fell within the nebula so that 
we have only used the Nod 2 measurement. By equating the continuum measured at
19$\mu$m in the LH measurement with the same continuum measured in the SH Nod 2
we obtain a ratio of LH/SH=2.5.  The ratio of SH to SL was obtained by equating
the \ion{[Ne}{ii]} fluxes in the two measurements. All fluxes were measured 
using the Gaussian line-fitting routine.The measured emission line intensities 
are given in Table 1, after correcting the SH measurements by the factor 2.5 
and the SL measurements by a factor of 4.07. The H$\beta$ flux found from the 
infrared hydrogen lines (especially the lines at 7.48$\mu$m and
12.37$\mu$m) using the theoretical ratios of \citep{hummer},
is 1.15x10$^{-11}$erg~cm$^{-2}$~s$^{-1}$, which is only slighly smaller than 
the H$\beta$ found from the radio flux density. This indicates that the LH 
diaphragm measured almost the entire nebula.

\subsection{Visual spectrum}

There are two measurements of the visual spectrum of He2-108, probably because 
it is weak and not visible to northern observatories. The measurements by
\cite{tpp} are considered by these authors to be 
less accurate than measurements of other PNe they have made. An accuracy of 
about 30\% is given by these authors.
The visual spectrum has also been measured in the 1990 Acker-Stenholm ESO 
survey of southern PNe. The full results of this survey have not yet been
published, but Acker has sent us the reduced spectrum of this object. The 
results are given in Table 11. The values are corrected for extinction; those 
in col.3 by the authors and those in col.4 by us in an attempt to produce the
expected Balmer decrement. The differences in the blue part of the spectrum 
reflects uncertainty of the measurements.

\begin{table}[h]
\caption[]{Visual spectrum of He2-108.}
\begin{center}
\begin{tabular}{llccc}
\hline
\hline
\multicolumn{1}{c}{$\lambda$} & Ion & \multicolumn{2}{c}{Intensities$^{\dagger}$}& Average\\ \cline{3-4}
\multicolumn{1}{c}{(\AA)}& & (1) & (2)  & Intens.\\
\hline

3727  & \ion{[O}{ii]} & 138   &     &  138 \\
3869  & \ion{[Ne}{iii]} & 4.7 & 20:  &  5: \\
4340 & \ion{H$\gamma$}& 43   & 44  & 43    \\
4363 & \ion{[O}{iii]} & 1.1 &      &  1.1: \\
4471 & \ion{He}{i}    & 6.8  &     &  6.8 \\
4861 & \ion{H$\beta$} & 100 & 100 & 100 \\
5007 & \ion{[O}{iii]} & 200 & 167   & 190 \\
5876 & \ion{He}{i}    & 17   & 17.5  & 17.5 \\
6563 & \ion{H$\alpha$}& 295  & 310   & 300      \\
6584 & \ion{[N}{ii]}  & 105  & 90  & 100 \\
7135 & \ion{[Ar}{iii]}&      & 15   &  15  \\
C(H$\beta$)    &        & 0.4 & 0.5  &    \\
\hline
\end{tabular}
\end{center}

$^{\dagger}$ References: (1) \cite{tpp}, (2) 
Acker (priv.comm.).\\ 
(:) indicates uncertain values.\\ 
C is the extinction (see text)
\end{table}

\subsection{The ultraviolet spectrum }

There are 6 IUE spectra of He2-108: 4 low resolution shortwave spectra and 2
low resolution longwave spectra. The spectra are dominated by the bright 
central star. Our interest was to look for evidence for nebular emission in the
carbon ions, either \ion{[C}{iii]} $\lambda$1907 \AA~or \ion{[C}{ii]}
$\lambda$2325 \AA. No \ion{[C}{ii]} emission could be seen. A
\ion{[C}{iii]} $\lambda$1907 \AA~line is seen in one of the four
shortwave spectra: SWP14181. Strangely this line is not seen in the other three spectra whose exposure time is longer than
SWP14181. The observed line is well above the noise and is at the correct 
wavelength, further the position of all four spectra are the same and only the
position angle is different. Until new observations are available we will 
regard this measurement as an upper limit to the intensity. The observed value
is 3 x 10$^{-13}$erg~cm$^{-2}$~s$^{-1}$, which after correction for extinction
becomes  3.3 x 10$^{-12}$erg~cm$^{-2}$~s$^{-1}$. There are no other obvious
nebular lines in these spectra.

\subsection{Electron density}

There is only a single ion where a line ratio can be used to determine 
$N_{\mathrm{e}}$. This is the \ion{[S}{iii]} ratio 33.4/18.7 in the infrared.
The ratio measured, 0.485, leads to a density of 2200 cm$^{-3}$. The ratio of
the \ion{[Ar}{iii]} lines in the infrared, 8.99/21.8, is not very dependent on
the density in this range but it is consistent with this value of density. The
abundances are not dependent on the density in this range.

\subsection{Electron temperature}

Only three ions have lines originating from energy levels far enough
apart that their ratio is sensitive to the electron temperature. These
are listed in Table 5. The temperature found from the \ion{[O}{iii]} lines is 
uncertain. A somewhat lower value of 8\,730 is given by \cite{mkenna} but the authors do not give the details of the spectra. We will use a 
temperature of 8\,500 K with an uncertainty of about 500 K. 

\subsection{Ionic and element abundances}

The ionic abundances have been determined using equation 1 above with an
electron temperature of 8\,500 K and a density of 2\,200 cm$^{-3}$.
The results are given in Table 12, where the columns are arranged as in Table 
12. The ICF is usually 1, except for nitrogen where it is assumed that the
ratio N$^{+}$/N$^{++}$= O$^{+}$/O$^{++}$. The principal ionization stage has
been measured in iron, argon and carbon but a small correction has been made 
for the singly ionized state. For argon, the similarity of the ionization 
potentials to nitrogen is the basis for the ICF used.

\begin{table}[htbp] 
\caption[]
{Ionic concentrations and chemical abundances in He2-108.
Wavelength in Angstrom for all values of $\lambda$ above 1000, otherwise
in $\mu$m.}

\begin{center}

\begin{tabular}{lccccc}
\hline
\hline
Ion & $\lambda$ & Int./H$\beta$ &  $N_{\mathrm{ion}}$/$N_{\mathrm{p}}$ 
& ICF & $N_{\mathrm{el.}}$/$N_{\mathrm{p}}$\\
\hline
He$^{+}$  & 5875 & 17 &  0.11 &    & 0.11    \\
C$^{++}$ & 1909 & $\leq$18 & $\leq$1.7(-4) & 1.1   & $\leq$1.9(-4) \\
N$^{+}$   & 6584 & 105 &  3.2(-5) & 1.9    & 6.0(-5)  \\
O$^{+}$  & 3727 & 67 &  1.4(-4)  &      &     \\
O$^{++}$  & 5007 & 200 & 1.35(-5) & 1  & 2.8(-4) \\
Ne$^{+}$  & 12.8 & 102  &  1.72(-4) &     &    \\
Ne$^{++}$ & 15.5 & 16.5 &  1.14(-5)  &  1  & 2.9(-4)  \\ 
S$^{+}$   & 6731 & 2 &  1.4(-7)  &     &     \\
S$^{++}$  & 18.7  & 72.2 &  7.46(-6) &    &    \\
S$^{+3}$  & 10.5  & 3.6  &  4.4(-7) & 1 & 8.1(-6) \\
Ar$^{++}$ & 8.99  & 25 &  2.7(-6) & 1.9   & 5.1(-6)   \\
Fe$^{++}$ & 22.9 & 1.02  & 3.6(-7) & 1.2  & 4.0(-7) \\    

\hline
\end{tabular}
\end{center}
Intensities given with respect to H$\beta$=100.

\end{table}

\section{NGC\,6629}

NGC\,6629 (PN G009.4-5.0, IRAS 18226-2313) is classified as an elliptical,
almost round nebula. It is slightly larger than Tc1 and He2-108, but smaller 
than NGC\,1535. Its size down to the 10\% level is given by \cite{tylenda} as 16.6\arcsec x 15.5\arcsec. It is surrounded by a halo which has a 
diameter of about 40\arcsec and is more compressed on the north side. The 6 cm
radio flux density is given by \cite{griffith} as 277 mJy.
\cite{ma1} give a 6 cm flux density of 292 mJy while \cite{ma2} find a 2 cm flux density of 234 mJy, which corresponds to a
value of 260 mJy at 6 cm. We will use a value of 270 mJy for the 6 cm radio 
flux density which predicts a value of H$\beta$=9.3 x 10$^{-11}$erg~cm$^{-2}$~s$^{-1}$ for the values of electron temperature and helium abundance given below.
Since the observed H$\beta$=1.18 x 10$^{-11}$erg~cm$^{-2}$~s$^{-1}$ the 
extinction constant C=0.896 or E$_{B-V}$=0.61.

\subsection{Infrared Spectrum}

The IRS measurement of NGC\,6629 was centered at RA(2000) 18$^{h}$25$^{m}$42.5$^{s}$ and Dec(2000) -23\degr12\arcmin10.1\arcsec. This is  very close to 
the center measured by \citep{kerber} of RA(2000)
18$^{h}$25$^{m}$42.45$^{s}$ and Dec(2000) -23\degr12\arcmin10.59\arcsec, which 
is presumably the coordinate of the central star. Because the nebula has a size
close to the sixe of the large LH diaphragm most of the nebula was within the 
LH diaphragm. The SH diaphragm measured only part of the nebula.  By equating 
the continuum measured at 19$\mu$m in the LH measurement with the same 
continuum measured in the SH diaphragm we obtain a ratio of LH/SH=2.7. This 
number is somewhat uncertain because the spectrum is rather noisy. This ratio can also be obtained by comparing both the \ion{[Ne}{iii]} 15.5/36.0 line ratio
and the \ion{[Ar}{iii]} 21.8/8.99 line ratio since both of these ratios have 
only a small dependence in electron density and temperature. We obtain an LH/SH
ratio of 1.71 from the \ion{[Ne}{iii]} lines and a value of 2.44 from the
\ion{[Ar}{iii]} lines. An average value of LH/SH=2.1 was used. The ratio of 
SH to SL was obtained by equating the \ion{[Ne}{ii]} the \ion{[S}{iv]} fluxes 
in the two measurements which leads to SH/SL=1.10. All fluxes were measured 
using the Gaussian line-fitting routine.The measured emission line intensities 
are given in Table 1, after correcting the SH measurements by the factor 2.1 
and the SL measurements by a factor of 2.3. The H$\beta$ flux found from the 
infrared hydrogen lines (especially the lines at 7.48$\mu$m and
12.37$\mu$m) using the theoretical ratios of \cite{hummer}
at a temperature of 8600 K is 8.45x10$^{-11}$erg~cm$^{-2}$~s$^{-1}$, which is 
only slighly smaller than the H$\beta$ found from the radio flux density. This 
indicates that the LH diaphragm measured most of the nebula.

\subsection{Visual spectrum}

There are several measurements of the visual spectrum of NGC\,6629 in the
literature. The two measurements quoted here are probably the best. These are
those of \cite{milingo3} and \cite{aller2} and 
are given in Table 13. In addition \cite{kings2} have 
measured the \ion{[O}{ii]} ratio 3726/3729 to be 1.63$\pm$0.50 and the 
\ion{[S}{ii]} ratio 6717/6731 to be 0.74$\pm$0.04.

The extinction coeficients, C, found from the Balmer decrement and given at the bottom of the table, are essentially the same as that found from the radio flux
density (given at the beginning of this section).

\begin{table}[h]
\caption[]{Visual spectrum of NGC\,6629.}
\begin{center}
\begin{tabular}{llccc}
\hline
\hline
\multicolumn{1}{c}{$\lambda$} & Ion & \multicolumn{2}{c}{Intensities$^{\dagger}$}& Average\\ \cline{3-4}
\multicolumn{1}{c}{(\AA)}& & (1) & (2)  & Intens.\\
\hline

3727  & \ion{[O}{ii]} & 36.2   & 41.1  &  38 \\
3869  & \ion{[Ne}{iii]} & 40.8 & 33.3  &  37 \\
4340 & \ion{H$\gamma$}& 46.3   & 46.5  & 46.4  \\
4363 & \ion{[O}{iii]} & 2.7 & 2.8  &  2.75 \\
4471 & \ion{He}{i}    & 4.6  & 4.43  & 4.5 \\
4861 & \ion{H$\beta$} & 100 & 100 & 100 \\
5007 & \ion{[O}{iii]} & 670 & 674   & 672 \\
5517 & \ion{[Cl}{iii]} & 0.45    &     & 0.45    \\
5538 & \ion{[Cl}{iii]} & 0.4   &     & 0.4 \\
5876 & \ion{He}{i}    & 15.2  & 12.0  & 14 \\
6312 & \ion{[S}{iii]} &  0.5  &      &   0.5 \\
6563 & \ion{H$\alpha$}& 286  & 289  & 287      \\
6584 & \ion{[N}{ii]}  & 10.8  & 11.0  & 10.9 \\
6717 & \ion{[S}{ii]}  &  0.6    & 0.53   & 0.53  \\
6731 & \ion{[S}{ii]}  &  0.7    & 0.71    & 0.71  \\
7135 & \ion{[Ar}{iii]}& 12.7  & 12.1  & 12.4  \\
9532 & \ion{[S}{iii]}  & 29.2   &     & 29.2 \\
C(H$\beta$)    &        & 0.8 & 0.96  &    \\
\hline
\end{tabular}
\end{center}

$^{\dagger}$ References; (1) \cite{milingo3}, (2) 
\cite{aller2}\\ 
(:) indicates uncertain values.\\ 
C is the extinction (see text)
\end{table}

\subsection{The ultraviolet spectrum }

There are seven IUE spectra of NGC\,6629: three low resolution shortwave spectra and four
low resolution longwave spectra. The diaphragm was centered near the edge of 
the nebula for some of the spectra and closer to the central star for other 
spectra. All spectra appear to be dominated by the bright central star. This 
indicates that much of the nebula is within the diaphragm but it is difficult
to specify exactly how much of the nebula is being measured. The only clearly 
nebular emission is the \ion{[C}{iii]} $\lambda$1907 \AA~line which is seen in 
all three shortwave spectra. It has the same intensity in all spectra: 2.5 x
10$^{-13}$erg~cm$^{-2}$~s$^{-1}$. Corrected for extinction this becomes 2.2 x
10$^{-11}$erg~cm$^{-2}$~s$^{-1}$; thus the ratio of the line to H$\beta$ is
23.7 (when H$\beta$=100). This value will be used when determining the carbon 
abundance but it is a lower limit since some of the nebular emission may be
outside the nebula. Possible \ion{[C}{ii]} $\lambda$2325 \AA~cannot be seen.

\subsection{Electron density}

The ions used to determine $N_{\mathrm{e}}$ are listed Table 4.
The electron density appears to be about 2000 cm$^{-3}$.

\subsection{Electron temperature}

Four ions have lines originating from energy levels far enough apart that their
ratio is sensitive to the electron temperature. These are listed in Table 5. An electron temperature of 8\,700 K is 
found with an uncertainty of less than 300 K. No temperature gradient is
apparent.

\subsection{Ionic and element abundances}

The ionic abundances have been determined using  equation 1 above with an
electron temperature of 8\,700 K and a density of 2\,000 cm$^{-3}$. The results
are given in Table 14, where, as in Table 6, the first column lists the ion
concerned, the second column the line used for the abundance determination and
the third column gives the intensity of the line used relative to H$\beta$=100.
The fourth column shows the ionic abundances, and the fifth column gives the 
Ionization Correction Factor (ICF), determined empirically. In all cases but 
one, when the ICF is greater than 1, the principal ionization stage of that 
element has been observed. The single exception is nitrogen, where it is
assumed that N$^{+}$/N$^{++}$=O$^{+}$ /O$^{++}$.

\begin{table}[htbp] 
\caption[]
{Ionic concentrations and chemical abundances in NGC\,6629.
Wavelength in Angstrom for all values of $\lambda$ above 1000, otherwise
in $\mu$m.}

\begin{center}

\begin{tabular}{lccccc}
\hline
\hline
Ion & $\lambda$ & Int./H$\beta$ &  $N_{\mathrm{ion}}$/$N_{\mathrm{p}}$ 
& ICF & $N_{\mathrm{el.}}$/$N_{\mathrm{p}}$\\
\hline
He$^{+}$  & 5875 & 14.0 &  0.094 &  ?  & 0.096    \\
C$^{++}$ & 1909 & 23.7 & 1.9(-4) & 1.1   & 2.1(-4)   \\
N$^{+}$   & 6584 & 10.9 &  3.23(-6) & 14  & 4.5(-5)    \\
O$^{+}$  & 3727 & 38 &  3.5(-5)  &      &     \\
O$^{++}$  & 5007 & 672 &  4.4(-4) & 1.1  & 4.8(-4) \\
Ne$^{+}$  & 12.8 & 9.95  &  1.8(-5) &     &    \\
Ne$^{++}$ & 15.5 & 89.1 &  6.06(-5)  &    &      \\ 
Ne$^{++}$ & 3869 & 37  &  7.2(-5)  &  1  & 8.4(-5)  \\ 
S$^{+}$   & 6731 & 0.7 &  5.3(-8)  &     &     \\
S$^{++}$  & 18.7  & 18.2 &  1.77(-6) &    &    \\
S$^{+3}$  & 10.5  & 14.4  &  3.9(-7) & 1 & 2.2(-6) \\
Ar$^{++}$ & 8.99  & 12.7 &  1.35(-6) &    &    \\
Ar$^{++}$ & 7135 & 12.4 &  1.67(-7) &  1.2  & 2(-6)   \\
Cl$^{++}$ & 5538 & 0.4 &  9.7(-8) &  1.2:  & 1.2(-8):    \\
P$^{++}$ & 17.9 & 0.61  &  4.5(-8) & 1.2:  & 2.0(-7):    \\ 

\hline
\end{tabular}
\end{center}
Intensities given with respect to H$\beta$=100.

\end{table}

\subsection{Errors}

We refer here to possible abundance errors in all of the PNe studied here. This
is difficult to specify because there are errors due to the measurements, the
electron temperature and the ICF. There is only a neglible error due to 
uncertainties in the electron density. The measurement error depends on the
strength of the line; for the stronger lines it is probably less than 15\%. The
values of $T_{\mathrm{e}}$ appear to be independent of the ionization 
potential in all
PNe considered here. For at least two of the nebulae the uncertainty could be 
as large as 1000 K. The temperature uncertainty plays only a small role for 
the infrared lines but is much more important for the ultraviolet lines. Taken
together we estimate that for all elements except carbon the abundance 
uncertainties are not more than 20-30\% for those elements for the ICF is
close to unity. When the ICF is higher than 1.5 the abundance uncertaintyies 
are about 50\%. For carbon the ICF is usually unity but the abundance is very
temperature dependent. The error is probably slightly higher, of the order of 
50\% for this element. The largest error for helium occurs in the PNe with
low temperature central stars where neutral helium is present. This occurs in 
Tc1, but may also occur in NGC\,6629 and He2-108. Other sources of erroe for 
the helium abundance are probably small.

\section{Model}

In order to obtain as nearly a correct model as possible, the star as
well as the nebula must be considered.
Modeling the nebula-star complex will allow characterizing not only
the central star's temperature but other stellar parameters as well
(ie., log g and luminosity). It can determine distance and other
nebular properties, especially the composition, including the
composition of elements that are represented by a single stage of
ionization, which cannot be determined by the simplified analysis
above.  This method can take the presence of dust and molecules into account 
in the nebular material, when there is any there, making it a very
comprehensive approach.  While the line ratio method is simple and
fast, the ICFs rest on uncertain physics.  To this end, modeling
serves as an effective means, and the whole set of parameters are
determined in a unified way, assuring self consistency.  Also, in this
way one gets good physical insight into the PN, the method and the
observations. Thus, modeling is a good approach to an end-to-end
solution to the problem. We use various models in Cloudy's library.

\subsection{Tc 1} 

\subsubsection{Assumptions} 

\cite{tylenda} give a diameter of  
12.9\arcsec ~$\times$ 12.2\arcsec~ for this PN.  We have used a diameter of 
12.55\arcsec~ in our modeling.  

\begin{table}[t] 
\caption[]  
{Parameters representing the final model for Tc 1.} 
\begin{tabular}{ll}  
\hline\hline  
Parameter & Value\\   
\hline 
\hline 
{\it Ionizing source} &\\ 
{\it Model atmosphere} & WMbasic \\ 
$T_{eff}  $ &34,700 K \\ 
Log g    & 3.30 \\ 
Log z    & -0.3 \\ 
Luminosity & 1480 $L_{\smallsun}$\\ 
\\ 
{\it Nebula}    & \\ 
\\Density profile & constant density 2850/cc \\ 

Abundance & H     \hskip 1.0cm He    \hskip 0.9cm C    \hskip  
1.0cm N \\ 
      & 12.000\hskip 0.3cm 10.916\hskip 0.6cm 8.674\hskip  
0.4cm 7.590 \\ 
      & O     \hskip 1.0cm Ne    \hskip 0.9cm Mg   \hskip  
0.9cm Si \\ 
      & 8.431 \hskip 0.4cm 7.481\hskip 0.4cm 5.5 \hskip  
0.5cm 5.778\\ 
      &P     \hskip 0.9cm  S     \hskip 1.0cm Cl    \hskip 0.9cm Ar  \hskip  
0.8cm  \\ 
      & 5.3\hskip 0.4cm 6.203 \hskip 0.4cm 4.963\hskip 0.6cm 6.478 \hskip  
0.4cm\\ 
Size      &$6.275\arcsec$ (radius)\\ 
Distance  & 1.80~kpc\\ 
Dust grains &  Graphites of single size $1.0\mu$m; \\ 
inner radius & 1.077e16cm  \\ 
outer radius & 1.690e17cm \\ 
Filling factor & 1.0\\ 
\hline 
\hline 
\end{tabular} 
\end{table} 

\subsubsection{Model results} 

Numerous models were run and we found that there was the primary problem 
of fixing the stellar effective temperature.  While the ionization of carbon,  
neon and argon indicated a somewhat lower $T_{eff}$, the observed  
\ion{[O}{iii]} lines required higher value. We tried a range of temperatures,  
distances, densities, density profiles and various model atmospheres.  Some 
models have been tried with stellar wind and some without while some with 
simple black body atmospheres for the CSPN. Observed \ion{[O}{iii]} line fluxes
seem to be unusually high and in trying to match them, Ne and Ar moved up to 
higher stages of ionization.  Many observed lines would suggest a cooler $T_{eff}$ than what our final model indicated.  In the final model, shown in Table 15,
we have used a low metallicity model atmosphere mainly to take care of infusing
more photons in the wavelength region below 912~\AA~without increasing $T_{eff}$. The metallicity is only half that of Sun. The value of gravity was also
kept as low as possible for a similar boost in the input stellar radiation.   
We note that the galactic latitude of the CSPN is only around -9 degrees and 
the fact that we were forced to use a low metallicity model atmosphere for such
a low latitude object shows the extreme to which a modeler is driven, when 
faced with anomalous nebular line emission.   

\begin{table*}[t]
\caption[]
{The emission line fluxes ($H\beta=100$) for Tc 1}
\label{tab-14}
\begin{tabular}{rrrrrrrrrr}
\hline\hline

Label & Line$^{\dagger}$ & Model flux & Obsd. flux  &&&  Label & line & Model flux & Obsd. flux \\
    &      &            & (dereddened)&&&       &      &   &  (dereddened)\\
\hline
TOTL & 4861A &  100.00	&  100.00	&&&  S  3 & 6312A &    0.40	&    0.46 \\
C  2 & 1335A &   10.35	&   11.43	&&&  O  1 & 6363A &    0.78	&    0.05 \\
C  3 & 1478A &    0.00	&    6.65	&&&  N  2 & 6548A &   32.43	&   30.10 \\
C  1 & 1561A &    0.34	&    1.26	&&&  N  2 & 6584A &   95.70	&   93.20 \\
C  2 & 1761A &    0.18	&    2.08	&&&  S II & 6716A &    2.38	&    2.15 \\
C  3 & 1907A &   16.56	&   14.76	&&&  S II & 6731A &    3.70	&    3.41 \\
C  3 & 1910A &   11.85	&   11.94	&&&  Ar 3 & 7135A &   15.52	&    6.99 \\
TOTL & 2326A &   86.80	&   44.59	&&&  O II & 7323A &    9.04	&    5.43 \\
O II & 3726A &  171.42	&  131.10	&&&  O II & 7332A &    7.23	&    4.57 \\
O II & 3729A &   91.40	&   86.59	&&&  Ar 3 & 7751A &    3.74	&    1.66 \\
S II & 4070A &    0.84	&    0.62	&&&  Fe 2 & 8617A &    0.02	&    0.01 \\
S II & 4078A &    0.27	&    0.18	&&&  C  1 & 8727A &    0.05	&    0.01 \\
Fe 2 & 4244A &    0.00	&    0.02	&&&  S  3 & 9069A &    7.64	&   12.51 \\
Fe 2 & 4359A &    0.00	&    0.01	&&&  TOTL & 9850A &    0.54	&    0.04 \\
TOTL & 4363A &    0.53	&    0.56	&&&  Ar 2 & 6.980m &   14.12 &   32.84 \\
P  2 & 4669A &    0.02	&    0.01	&&&  Ar 3 & 9.000m &   16.23 &    6.49 \\
O  3 & 4959A &   40.50	&   41.71	&&&  S  4 & 10.51m &    0.51 &   0.47 \\
O  3 & 5007A &  121.92	&  123.00	&&&  Ne 2 & 12.81m &   19.68 &  37.41 \\
Ar 3 & 5192A &    0.08	&    0.03	&&&  Ne 3 & 15.55m &    3.24 &  1.46 \\
N  1 & 5198A &    0.16	&    0.02	&&&  P  3 & 17.89m &    0.72 &  0.73 \\
N  1 & 5200A &    0.06	&    0.02	&&&  S  3 & 18.67m &    12.47&  14.03 \\
Cl 3 & 5518A &    0.27	&    0.28	&&&  Ar 3 & 21.83m &    1.05 &   0.43 \\
Cl 3 & 5538A &    0.29	&    0.30	&&&  Fe 3 & 22.92m &    0.27 &   0.29 \\
O  1 & 5577A &    0.04	&    0.00	&&&  S  3 & 33.47m &    4.77 &   6.21 \\
N  2 & 5755A &    1.23	&    1.08	&&&  Si 2 & 34.81m &    0.32 &   0.30 \\
O  1 & 6300A &    2.45	&    0.12	&&&     &        &          & \\
\hline
\end{tabular}

Absolute $H\beta$ flux Model: $5.83\times10^{-11}ergs~cm^{-2}~s^{-1}$
Obsn: $6.15\times10^{-11}ergs~cm^{-2}~s^{-1}$ \\
$\bf{Notes}$: `A' in col.  `Line' signifies Angstrom;
`m' signifies $\mu$m.  In col. `Label', we have
followed the notation used by Cloudy for atoms and ions; this will make
identifying a line in Cloudy's huge line list easy.  Neutral state
is indicated by `1' and singly ionized state by `2' etc., `TOTL' typically
means the sum of all the lines in the doublet/multiplet; or it could
mean sum of all processes: recombination, collisional excitation,
and charge transfer.
Some elements are represented by usual notation as per Cloudy.

\end{table*}

At this stage we were not sure whether any extra source of energy was present, 
as there was no observational clue, but the above facts forced us to look at 
all different possibilities.  It might as well be that model atmospheres do not
realistically represent the stellar radiation.  Another fact is that the 
accuracy of the optical spectra containing the \ion{O}{iii}  
lines is quite high, an error of only 8\% is quoted by the authors for the  
spectrophotometry.  Therefore we surmise that something very interesting is  
happening in the formation of \ion{O}{iii} lines but are unable to hazard any  
guess.  But while we did these adjustments in modeling, lines of neon, argon
and sulphur gave trouble.  The sulphur lines were improved to get a better fit 
by adjusting the DR~(dielectronic recombination) rates since the DR rates are 
poorly known.  The final model output spectra are presented in table 16, where
it is clear that the model fluxes for most low ionization stages (\ion{O}{ii},
\ion{C}{ii}, and \ion{Ne}{ii}) are too high.

To reproduce the observed IR dust continua, we found that graphites gave a 
better fit and used them in our modeling rather than silicates. The grains 
included in the modeling were the Cloudy's set called Orion distribution which 
has a bias towards larger grain sizes. The match to the observed IR continua is
reasonable (see Fig.1). So the final model we present is the best we could 
achieve. Overall we feel our exercise raises more pertinent questions than 
answers as this PN seems to throw lots of challenges to our current 
understanding of 
nebular physics.  We are inclined to recommend the abundances as determined by 
the ICF method for this PN. A very pertinent point we want to highlight is the 
fact that this PN shows nebular absorption lines too; see \cite{williams} and to the best of our knowledge existing photoionization codes are 
yet to have a provision for computing the equivalent widths of such lines so 
that they can also be compared with observed values.  This would make the 
criteria for a good fit (with observation) more stringent.  Presently all 
models published till date used only nebular emission lines alone. Incorporatingthe formation of absorption lines would be the next major paradigm shift in 
photoionization modeling.

\subsection{Modeling NGC 1535}

We now describe our unsuccessful efforts to make a photoionization model for 
this PN. It is necessary to go into the details since it reveals insights 
into aspects of nebular physics which are normally taken for granted as well 
known.  

\subsubsection{Assumptions}

The appearance of NGC\,1535 from an image taken by \cite {schwarz} is nearly spherical.  The IR and radio measurements described earlier give an
absolute $H_{\beta}$ flux that is consistent with a diameter of 45\arcsec, 
that includes a low density outer zone beyond a high density innerzone of 
diameter of 19\arcsec. We wanted to include a density profile for the nebula to
describe the variation of the number density N(H) with the radius and so 
derived a template profile from the $H_\alpha$ image taken from http://astro.uni-tuebingen.de/groups/pn/. The image was taken on an as-is basis and imported 
into the IRAF and using cross-cuts at different azimuths an average profile was
generated and then normalized to have a peak value of 1300/cc. The nebular 
radius was normalized to 22.5\arcsec, to include the low density region as 
well. We used this profile as a starting value but later have experimented with
modifications to it as well as tried simple constant density models. Though the
presence of dust and $H_2$ molecules is observed in this PN we did not include 
them in our simulations.

\subsubsection{Modeling and its implications}

Though we had tried a series of numerical model computations for this object by
choosing various options in the parameter space, the primary road-block we 
faced was this: the observed fluxes of \ion{[O}{iv]} 25.88$\mu$m and the \ion{He}{ii} 1640\AA~were too high to reproduce while those of specie \ion{O}{ii}, \ion{S}{ii} and \ion{Cl}{iii} were too low to be reproduced by the models.
Many years ago, \cite{aller2}  modelled this object and he modified the 
incident energy spectrum to get a good fit.  Later 
\cite{ak} (hereafter AK) had to introduce energy from stellar wind at a $T_e$ of $3\times10^5$ K to adequately reproduce the observations.  While Aller's adoptation is 
purely ad hoc, that of AK is not viable physics-wise.  AK had predicted that 
such an extra source of energy should make the PN an x-ray emitter but this is
not the case as shown by \cite{guerrero}.  Secondly they claim
that this additional energy source produces a good model matching observations.
But looking at their final results, (see table 13 of their paper) we find that 
they did not match the diagnostic lines of \ion{O}{ii} and \ion{S}{ii} well.  
They did not include the optical lines of \ion{[Cl}{iii]}, \ion{He}{ii} 1640\AA~
and \ion{[O}{iv]} 25.88$\mu$m among others, whereas we used a comprehensive set 
of multi-wavelength spectral observations in this work. A more important point 
is that they claim that a proper nebular model with rigourous radiative 
transfer would match the observations very well.  On the other hand, we have 
tried models having proper windy model atmospheres (for the CSPN) with Cloudy 
(wherein radiative transfer is handled by escape probability), and it did not 
work out well.  We have even tried to infuse additional photons below 226A in 
the input stellar radiation but this did not work out.  It is clear that AK's
suggestion of wind plasma as the extra energy source is ruled out. 

As mentioned above \ion{He}{ii} and \ion{[O}{iv]} lines are very strong and to 
produce them we needed a model atmosphere with a $T_{eff}$ of 120,000 K.  But 
all other lines then do not match properly with such a model.  We experimented 
with a whole range of values of the CSPN parameters but did not succeed. We 
note that from FUV observations of the continua of the CSPN by FUSE, \cite{herald} (HB hereafter) obtain a $T_{eff}$ of only 66,000 K.  When
we ran a model with the CSPN parmeters as given by HB, it did not reproduce the
nebular spectra well.  Diagnostic lines did not match; \ion{He}{ii} and \ion{[O}{iv]} lines were weak.  More importantly HB give a luminosity of around 4000 
L\smallsun~ which when used in our model, gave the transmitted flux as nearly 
80\% of the incident flux.  We found that a luminosity of 550 L\smallsun~would 
be sufficient to produce the absolute ${H_\beta}$ flux correctly.  But only a 
high luminosity of 4000\smallsun~is compatible with the observed FUV continua 
of the CSPN.  This creates another problem because this PN is quite complex as 
it shows hydrogen molecular absorption lines in its FUSE spectrum. These $H_2$ 
molecules are attributed to be circum-nebular rather than interstellar by HB. 
It would be very difficult to imagine the presence of these molecules when the 
nebula leaks the incident stellar radiation enormously in the FUV and UV 
wavelengths!  They would simply be destroyed by such a strong radiation.  The 
dust content is not very high in this PN as shown by HB and so the idea of dust
grains shielding $H_2$ molecules from the strong radiation is also ruled out. 
In our modeling experiments we have even tried including a binary CSPN by way 
of introducing two sources of input radiation, a hot and a cooler star. We 
experimented with different combinations of temperature and luminosity but were
not able to reproduce the observations. 

In summary, the best we could intuitively guess was that since the complicated 
ionization strucuture as demanded by the observations of nebular emission lines
was impossible to reproduce, it is probable that the wind streaming into the 
nebular material is directly injecting highly ionized elements, particularly O 
and He.  We are led to think of such a scenario under the given context.  We 
feel that this PN is quite complicated as far as its physics and chemistry is 
concerned.  It is important that this new idea of `wind streaming' is 
observationally explored to establish its possibility, especially as it may be
applicable other PNe which are difficult to model, such as NGC\,2392. Shocks
have been suggested as the source of energy needed to increse the ionization in
NGC\,2392 and NGC\,1535 \citep[e.g.][]{PTL}. But if the shocks
result in a hot plasma as assumed by \cite{ak} they are in 
disagreement with the observations as discussed above. Thus the suggestion of
'ion streaming' is a good possibility for both NGC\,2392 and NGC\,1535.
We conclude that the determination of abundances is possible, at present, only 
by the ICF method for this PN.

\begin{figure} 
\centering 
\includegraphics[width=6.0cm, angle=270]{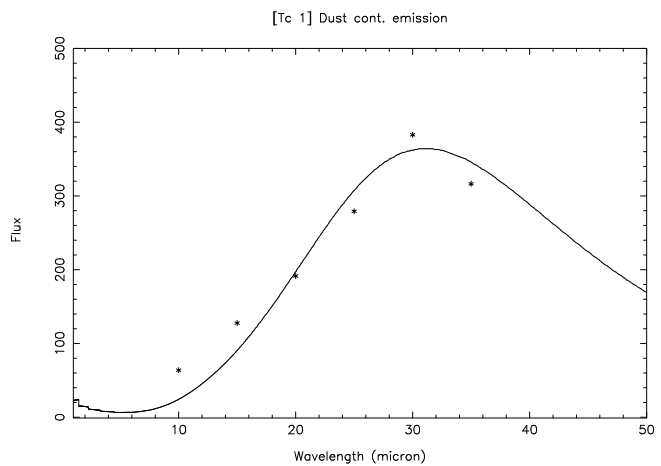} 
\caption{The IR dust continua of Tc~1. 
 The asterisks represent the observation from Spitzer 
and the continuous curve is the model output.} 
\label{fig-1} 
\end{figure}

\section{Discussion}

In Table 17 the abundances for the four PNe discussed above are summarized in 
the first four lines, followed by the abundances for five other nebulae which  
have already been determined. As mentioned in the introduction all these PNe 
have been discussed by \cite{mendez, mendez2} and form a very
homogeneous group. Not only are these nebulae excited by bright, rather low 
temperature central stars which are rather far from the galactic plane, the 
central stars all have spectra indicating that they are hydrogen rich. \cite{mendez3} classifies them all as either O(H) or Of(H).

As may be seen from the table,
the abundances of these PNe are rather uniform. The oxygen abundance varies by 
afactor of 1.9 but some of this may be caused by the known abundance gradient
with distance from the galactic plane. To better judge this effect, an 
approximate value of the distance of the PNe from the galactic center is given 
in col. 10 of the table. The nitrogen abundance, or better still the N/O ratio
also has the same low value for all the PNe, with the single exception of 
NGC\,2392. This low value of N/O is the same as found in the solar atmosphere,
which is also shown in Table 17. Also notice that the helium abundance of these
PNe is very similar to the solar value. Combining these abundances with the 
theoretical determinations of \cite{karakas} we obtain the following 
picture. These PNe originate from stars of low enough mass so that the second 
dredge-up or hot-bottom burning have not taken place. These processes would 
have increased both the 
helium abundance and the N/O ratio to values higher than observed. For 
comparison the last two entries in Table 17 show the abundances of PNe whose
central stars are of higher mass and have clearly undergone hot-bottom burning.

The ratio of carbon to oxygen (C/O) clearly varies for the PNe shown in Table 
17. The three lowest values are similar to the solar value, about 0.5. The 
highest C/O ratio, that of IC\,418 is almost four times as high. The PNe with 
the lowest C/O ratio probably originate in stars of similar mass to the sun 
while the higher C/O value originate in somewhat higher mass stars which have 
undergone the third dredge-up. Following \cite{karakas} the initial 
mass of the stars with the low carbon abundance is between 1 and 1.5M\smallsun,
while substantial carbon will be produced betweem 1.75 and 2.5M\smallsun. 
According to \cite{weide} the first group corresponds to a final
core mass of 0.55 to 0.57M\smallsun, while the group with substantial carbon 
will have a final mass of between 0.59 and 0.63M\smallsun.

\begin{table*}[!ht]
\caption[]{Elemental abundance of PNe with far-infrared data in addition to 
optical and UV data.}
\begin{center}
\begin{tabular}{|l| c c c c c c c  c| c c  |}
\hline
\hline
PNe & He/H & C/H & N/H & O/H & Ne/H & S/H & Ar/H & Cl/H & R$^\natural$ &  Ref.\\
& &$\times$10$^{-4}$&$\times$10$^{-4}$&$\times$10$^{-4}$&$\times$10$^{-4}$&$\times$10$^{-6}$&$\times$10$^{-6}$& $\times$10$^{-8}$ &         (kpc) &  \\
\hline
NGC\,1535 &  0.091    &  1.6   & 0.32  & 2.7   &  0.54   &   !.3  &  1.1  &  6.0  & 9.9  &   a  \\
Tc\,1     &$\geq$0.060 &  3.6   & 0.36  & 2.6   &  0.63   &   2.8  &  5.1  &  9.4  & 6.3  &   a  \\
He2-108   &  0.11     &$\leq$1.9& 0.60  & 2.8   &  2.9    &   8.1  &  5.1  &       & 4.1  &   a  \\
NGC\,6629 &  0.096    &  2.1   & 0.45  & 4.8   &  0.84   &   2.2  &  2.0  &  12:  & 6.0  &   a  \\
NGC\,6826 &  0.10     &  4.8   & 0.58  & 3.95  &  1.5    &   2.6  &  1.4  & 8.5   & 8.0  &   b   \\
IC\,418   &$\geq$0.072 &  6.2   & 0.95  & 3.5   &  0.88   &   4.4  &  1.8  & 12    & 8.8  &   c  \\
IC\.2448  &  0.094    &  2.7   & 0.55  & 2.5   &  0.64   &   2.0  &  1.2  &       & 8.0  &   d  \\
NGC\,2392 &  0.080    &  3.3   & 1.85  & 2.9   &  0.85   &   5.0  &  2.2  &  13   & 8.4  &   e  \\
NGC\,3242 &  0.092    &  1.95  & 1.0  & 3.8   &  0.90   &   2.8  &  1.7  &  7.0  & 8.1  &   f  \\
\hline
Solar     & (0.085)   & 2.7    & 0.675 & 4.9   & (0.84)  &  13.   & (2.5) & 31.  &  8.0  &   g  \\
\hline
NGC\,6302 &  0.17     & 0.6    & 2.9   & 2.3   &  2.2    &  7.8   &  6.0  & 34. &  6.4   &   h  \\
NGC\,6537 &  0.149    & 1.7    & 4.5   & 1.8   &  1.7    &  11.   &  4.1  & 24. &  6.0   &   j  \\

\hline

\end{tabular}
\end{center}
References: a) Present paper; b) Surendiranath \& Pottasch 2008, A\&A 483, 519;
c) Pottasch et al. 2004, A\&A 423, 593; d) Guiles, S. et al. 2007, ApJ 669, 
1282; e) Pottasch et al. 2008, A\&A 481, 393; f) Pottasch \& Bernard-Salas 
2008, A\&A 490, 715; g) Grevesse et al. 2010, ApSS 328, 179; h) Pottasch et al.
1999, A\&A 347, 975; j) Pottasch et al. 2000, A\&A 363, 767

\end{table*}

The observed nebular abundances do not permit a more quantitative determination
of the stellar masses than that given above. We can however, determine the
stellar masses which are predicted by stellar evolution theory to see if they
are consistent with the masses determined from the nebular abundances above. To
do this we make use of the summary of stellar evolution calculations given by
\cite{blocker} in the form of an HR diagram where the time of 
evolution from the AGB is marked on each evolutionary track (his Fig. 12). 
Lines of constant time (isochrones) are also shown in this figure. We have 
fixed the position of each of the central stars being discussed on this figure 
by determining the stellar temperature and age. The stellar temperature is 
taken from the spectra of the stars using the work of 
\cite{mendez,kudritzki2,pauldrach}, and is listed in Col.2 of Table 18.
In seven of the nine cases these temperatures are the same as are determined 
from the nebula \citep[e.g. see][]{pott7}. In two 
cases the spectroscopic temperature is lower: NGC2392, where the difference is
considerable, and NGC1535, where the difference is much smaller. The reason for
this difference is not yet understood, but is probably related to the heating 
of the nebula as described above.

The age is PN is determined from the observed size and expansion velocity of
the nebulae. These quantities have been taken from the values listed by
\cite{acker} and are given in Cols.5 and 6 of the table. When two 
values of velocity are listed by \cite{acker}, the value for the 
\ion{[N}{ii]} is used since this line is formed farthest out in the nebula. The
velocity is measured in the line-of-sight while the measured size is tangential
but because these nebulae are nearly round it is expected that these values may
be combined. Having determined the age of the nebula its position on the HR 
diagram is now fixed. The values of luminosity and core mass corresponding to 
this position are shown in cols.\,9 and 10 of Table 18. While uncertainties in 
the value of size and velocity may be considerable and the age determination 
only reliable to within 50\%, the values of luminosity and core mass found are 
much better determined because the isochrones are so closely spaced, i.e. the 
luminosity and core mass have a rather small dependance on the age in the low 
temperature range of the evolutionary tracks.

The core mass found in this way is the same as that deduced above from the
nebular abundances. Even in the prediction that the PNe with the high C/O 
ratios will have somewhat higher core masses appears to be fulfilled. We may 
say that the core masses predicted from the PNe abundances are consistent with 
those found from stellar evolution.

From the luminosity found (in col.9) and the assumption that the central star 
radiates as a blackbody with the temperature given in col.2 and the angular
radius found from the stellar magnitude and extinction listed in cols.3 and 4,
the distance can be computed. These distances are listed in the last column of
Table 18. In 6 of the 9 cases this distance agrees to within 15\% with the 
statistical distances given either by \cite{cahn} or \cite{stang2}. In the remaining three cases (NGC2392, NGC3242 and IC418) the
listed distance is about 60\% higher.

\begin{table*}[!ht]
\caption[]{Prediction of central star mass and distance from evolution theory}
\begin{center}
\begin{tabular}{|l| c c c c c c c  c| c c  |}
\hline
\hline
PNe &$T_{eff}$ &$m_{v}$ & C & Vel. & Rad. &  t &R$_{s}$/R\smallsun &L$_{s}$/L\smallsun & Mass & dist.\\
 &  K    &        &      & km/s &\arcsec &10$^{3}$sec &           &         &   M\smallsun  & kpc  \\
\hline
NGC1535 & 66\,000 & 12.11 & 0.08 & 20  & 10.5 & 5.3  & 0.543  & 5000 & 0.59 &  2.0 \\
Tc\,1   & 32\,000 & 11.38 & 0.36 & 12.5 & 5.0 & 5,4  & 2.17   & 4500 & 0.57 & 2.64 \\
He2-108 & 32\,000 & 12.72 & 0.53 & 12   & 5.5 & 9.2  & 1.96   & 3550 & 0.56 & 3.7  \\
NGC6629 & 46\,000 & 12.87 & 0.88 & 6.5  & 7.7 & 9.5  & 0.95   & 3800 & 0.57 & 1.8  \\
NGC6826 & 48\,000 & 10.68 & 0.07 & 11   & 12.7 & 8.4 & 0.935   & 4080 & 0.57 & 1.42 \\
IC\,418 & 36\,000 & 10.23 & 0.33 & 12   &  6   & 4.0 & 1.60   & 5000 & 0.60 & 1.25 \\
IC\,2448 & 65\,000 & 14.26 & 0.27 & 13.5 & 5   & 6.5 & 0.54   & 4550 & 0.58 & 4.05 \\
NGC2392  & 43\,000 & 10.63 & 0.22 & 53  & 22.4 & 1.7 & 1.55   & 7600 & 0.62 & 1.8  \\
NGC3242  & 75\,000 & 12.32 & 0.12 & 27.5 & 19  & 4.7 & 0.43   & 5100 & 0.59 & 1.75 \\

\hline

\end{tabular}
\end{center}

\end{table*}

\section{Conclusions}

With the help of Spitzer infrared spectra the abundances in four PNe have been
determined. These PNe are all excited by rather low temperature central stars
and have similar morphological and kinematic properties: they are all nearly
round and are rather far from the galactic plane. We are able to show that 
these nebulae have rather similar abundances of helium, oxygen, nitrogen, 
carbon and other elements. The resultant abundances are summarized in the first
four lines of Table 17. We then show that five other PNe with low temperature 
central whose abundances have been determined using Spitzer infrared spectra 
and have the same or similar morphological and kinematic properties also have 
the same or similar abundances.

By comparing these abundances with those predicted by nucleosynthesis models by
\cite{karakas} 
it is deduced that these PNe originate from stars of initial mass between 
1M\smallsun~and about 2.5M\smallsun, which according to \cite{weide}
correspond to core masses of between 0.56M\smallsun~and 0.63M\smallsun. These
values of core masses are compared with those determined from stellar evolution
theory using the observed temperature of the central star and the measured
age of the nebula. The core masses thus found are consistent with each other.
Two details reinforce this consistency. First, the higher core masses from the
evolutionary theory are found in PNe which have high C/O abundance
ratios as the models of \cite{karakas} predict. Secondly the distances
found from the stellar evolution are in general values expected from 
statistical distance scales.

There are a number of uncertainties which must still be considered. First, it 
is not understood why the central star temperature measured in NGC\,2392 is so 
much lower than that found from the nebula. Second, distances found by some
researchers are different than found here. For example, the distances given by
\cite{kudritzki2} for 6 of the PNe listed are 50\% higher than
we have found. This could be due to errors in their determination of the
stellar gravity from uncertain line profiles. Furthermore the expansion 
distances found for two of the nebulae (NGC\,3242 and IC\,2448) are at least a 
factor of 2 lower than we have found here. This should be carefully considered
in the future.

\begin{acknowledgements}

We duly acknowledge the use of SIMBAD and ADS in this research work. We have 
used the IUE spectra archive at the STSCI and we wish to thank the archive 
unit for the same. RS would like to acknowledge that a part of his research 
work was done when he was working at the Indian Inst. of Astrophysics, 
Bangalore. RS sincerely thanks his former colleagues J.S.Nathan and B.A.
Varghese for help with S/W upgradation.

\end{acknowledgements}

\bibliographystyle{aa}

\end{document}